\documentclass[12pt]{iopart}

\usepackage[active]{srcltx}

\newcommand{\be}{\begin{equation}}
\newcommand{\ee}{\end{equation}}
\newcommand{\bea}{\begin{eqnarray}}
\newcommand{\eea}{\end{eqnarray}}
\newcommand{\bd}{\begin{displaymath}}
\newcommand{\ed}{\end{displaymath}}

\newcommand{\qi}{ q^{-1}}

\newcommand{\ad }{a^{\dagger}}

\newcommand{\p}{ \partial_x }
\newcommand{\lb }{ \left( }
\newcommand{\rb }{ \right ) }

\begin{document}

\title[Symmetric Tamm-Dancoff $q$-oscillator]{Symmetric Tamm-Dancoff $q$-oscillator: representation,
quasi-Fibonacci nature, accidental degeneracy and coherent states}

\author{Won Sang Chung$^1$, A.M. Gavrilik$^2$, I.I. Kachurik$^{2,3}$, A.P. Rebesh$^2$}
\address{$^1$ Department of Physics
and Research Institute of Natural Science, College of Natural
Science, Gyeongsang National University, Jinju 660-701, Korea}
\address{$^2$  Bogolyubov Institute for
Theoretical Physics of NAS of Ukraine, 14b,  Metrolohichna Str.,
03680 Kyiv, Ukraine}
\address{$^3$  Khmelnytskyi National University, 11, Instytutska Str., 29016 Khmelnytskyi, Ukraine}
\eads{\mailto{mimip4444@hanmail.net}, \mailto{omgavr@bitp.kiev.ua}, \mailto{rebesh@bitp.kiev.ua}}

\begin{abstract}
In this paper we propose a symmetric $q$-deformed Tamm-Dancoff
(S-TD) oscillator algebra and study its representation, coordinate
realization, and main properties. In particular, the non-Fibonacci
(more exactly, quasi-Fibonacci) nature of S-TD oscillator is
established, the possibility of relating it to certain
$p,q$-deformed oscillator family shown, the occurrence of the
pairwise accidental degeneracy proven. We also find the coherent
state for the S-TD oscillator and show that it satisfies
completeness relation. Main advantage of the S-TD model over usual
Tamm-Dancoff oscillator is that due to ($q\leftrightarrow
q^{-1}$)-symmetry it admits not only real, but also complex
(phase-like) values of the deformation parameter $q$.
\end{abstract}

\pacs{02.20.Uw, 03.65.-w, 03.65.Fd, 05.30.Pr}
\maketitle

\section{Introduction}
Complex systems in diverse fields of quantum physics,
especially the systems with essential nonlinearities, can be
efficiently described by quantum algebras and deformed oscillator
algebras.
 In models of quantum field theory or particle theory, the involved
Bose-like deformed oscillators (implying unusual statistics of
particles and requiring modified commutation relations) can provide
better description of real quantum processes, in particular with
participation of pions, see \cite{Curado}, \cite{GavrSigma}.
The use of bosonic deformed oscillators for treating phonons
can significantly improve \cite{Monteiro} matching between
theory and experiment concerning the unstable phonon spectrum
 in ${}^4$He. There exist interesting application of deformed bosons picture for an effective
description of the spectra of excitons \cite{Harouni-09,Zeng-11}.
Some version of the $q$-deformed Bose gas model was applied to the
systems of interacting pointlike (structureless) particles in
\cite{Scarfone-09}. This result was recently extended in \cite{GM-13} where it
was shown that two
 factors of non-ideality or deviation from the ideal Bose gas picture:
(i) compositeness (two-fermionic or two-bosonic) of non-interacting Bose-like particles,
and (ii) the inter-particle interactions, can be jointly accounted for by using
definite model of deformed bosons.
 This extends the results of \cite{Scarfone-08} concerning effective description
 by $q$-bosons of the interacting (thus non-ideal) gas of bosons.

Among best known deformed oscillator models, there are such ones as
the Arik--Coon (AC) \cite{A-C76}, Biedenharn--Macfarlane (BM) \cite{Biedenharn,Macfarlane89},
and the $p,q$-deformed oscillator \cite{Chakrabarti91} models.
 Their popularity explains a diversity of existing applications of
 these models. Much less studied is so-called Tamm--Dancoff (TD)
$q$-deformed oscillator model \cite{Odaka,Chaturvedi,GR07}.
Note nevertheless that the TD-type $q$-deformed Bose gas was studied in \cite{GR-12,Algin-TD-13}.
Whereas AC model shows no energy level degeneracy, the TD model
possesses the  accidental pairwise degeneracy of levels. From the
three models, only BM oscillator admits not only real, but also
complex-valued deformation parameters.

In this paper, we introduce and study in some detail certain
extension of the TD  model, namely the ($q\leftrightarrow q^{-1}$)
symmetric Tamm-Dancoff (S-TD) oscillator whose main distinction and
advantage with respect to the TD model is the admittance of both
real and complex-valued deformation parameter $q$. Then, in the S-TD
or symmetrized modification of TD oscillator, the situation becomes
different when studying accidental degeneracy: there is no energy
level degeneracy for real $q$, but there appears an accidental
degeneracy when $q$ takes definite complex, phase-like, values of
$q$.

The plan of the paper is the following. After Introduction, in
 Sec. 1 we present the setup for the S-TD oscillator, its representation,
and the related S-TD (non-standard) version of $q$-calculus:
$q$-derivative, $q$-integral and the S-TD $q$-deformed exponent.
 In Sec. 2, we prove the non-Fibonacci, more concretely
quasi-Fibonacci properties of the S-TD $q$-oscillator.
 The next Sec. 3 is devoted to clarifying the issue of how to ``embed'' the S-TD
$q$-deformed model into certain modification of the two-parameter
$p,q$-family of deformed oscillator. The appearance of
accidental pairwise energy level degeneracy is analysed in Sec. 4.
In the last Sec. 5, the S-TD type deformed coherent states are
described, and their completeness is proven.
 The paper ends with Conclusions.

\section{S-TD $q$-oscillator and the related version of $q$-calculus}
To begin, we define the symmetric $q$-deformed bosonic Tamm-Dancoff
oscillator algebra:
\be \label{1.1}
aa^{\dagger} -   a^{\dagger}a=\frac{1}{2} ( 1 + (1-\qi )N ) )
q^{N} +  \frac{1}{2} ( 1 + (1-q) N) q^{-N},
\ee
\[
[ N, a^{\dagger} ] =
a^{\dagger}, \quad [N, a]= -a,
\]
 where
\be \label{1.2} \ad a = \{ N \} \equiv\frac{ N }{2} ( q^{ N-1} +
q^{-N+1} ) ,  \ee
 and $q$ is either real, \ $0< q \le \infty$, or complex phase-like: $q=\exp({\rm
 i}\theta)$, \
  $-\pi \le\theta\le
 \pi$.

 We can easily find the Fock-type representation of algebra (\ref{1.1}):
\bd N|n\rangle = n |n\rangle, ~~~~ n=0, 1, 2, \ldots , \ed \bd
a|n\rangle =
 \sqrt{\frac{ n }{2} ( q^{n-1} + q^{-n+1} )  }|n-1\rangle
  = \sqrt{\{ n \}}|n-1\rangle , \ed
\[
\ad |n\rangle =
\sqrt{\frac{ n+1 }{2} ( q^{ n} + q^{-n} )  }|n+1\rangle = \sqrt{\{
n+1\}}|n+1\rangle
\]
 where $\{ n \}$ denotes the S-TD type $q$-number or structure function
 defined as
\be \label{1.4} \{ n \} \equiv \varphi_{\tiny{STD}}(n) = \frac{n}{2}
( q^{n-1} +  q^{-n+1})
  = \frac {n}{2} q^{ - n+1} ( 1 + (q^2)^{n-1} ) .
\ee
 Remark that this S-TD $q$-bracket can be also written as a
``multiplicative hybrid'' of (the structure function
$\varphi(n)\!=\!n$ of) usual oscillator and of the BM oscillator
with the ($q\leftrightarrow q^{-1}$)-symmetric structure function
$\varphi_{BM}(n)\equiv[n]_q=\frac{q^n-q^{-n}}{q-q^{-1}}$, namely
\be
\{ n\}=n\,\frac{[n]_q\!-\![n\!-\!2]_q}{2} \qquad {\rm or} \qquad
 \{ n\}=n\,\frac{[2(n-1)]_q}{2\,[n-1]_q}.
\ee

\begin{center}
  \textbf{\emph{S-TD variant of $q$-calculus} }
\end{center}

In order to have a functional realization of this representation, we
consider the space $\cal{P}$ of all monomials in variable $x$, and
introduce its basis of monomials \be \label{1.5} |n \rangle
\leftrightarrow \frac{x^n}{ \sqrt{ \{n\} ! }}
    \vspace{-5mm}
\ee
 where
\[
\{n\} ! = \prod_{k=1}^n \{ k \} , ~~~ \{0\}! =1 .
\]
Then the functional realization of the algebra (\ref{1.1})-(\ref{1.2}) is given by
\[
 a \leftrightarrow D_x , ~~~~ \ad \leftrightarrow x , ~~~ N
\leftrightarrow x \p.
\]
 Here the new deformed derivative (different from well-known Jackson $q$-derivative
 \cite{Jackson}) is defined as
\be \label{1.8}
D_x \equiv \frac{ 1 }{ 2 } ( T_q + T_{q^{-1}} ) \p =  \frac{ 1
}{ 2 }  \lb q^{ x \p  } + q^{  -x \p } \rb \p ,
\ee
 where $T_q=q^{x \p }$ and such that $T_q f(x)= f(qx)$ .
  It is clear that $T_{q^{-1}}=T_q^{-1}$.

 The Leibnitz rule of the deformed derivative is then
given by
\[
D_x ( f(x) g(x) ) = ( D_x f ) (T_q^{-1} g )+( T_q f) (
D_x g ) -
\]
\[
\hspace{26mm} -\frac12 ( {\cal T }f ) ( T_q^{-1} \p g ) +
\frac12 ( T_q \p f ) ({\cal T}g),
\]
where
\[
{\cal T }f(x)=( T_q -T_q^{-1} )f(x)=
f(q x)- f(q^{-1} x ).
\]
 The deformed integral is defined as
\[
 \int Dx f(x) = 2 \int ( T_q + T_q^{-1} )^{-1} f(x) dx=
\]
\[
 \hspace{20mm} = 2 \sum_{n=0}^{\infty} (-1)^n \int dx f( q^{ 2n +1 } x )=
\]
\[
\hspace{20mm} = 2 \sum_{n=0}^{\infty} (-1)^n \int dx f( q^{-(2n+1) } x )=
\]
\be \label{1.13}
 \hspace{20mm} = \sum_{n=0}^{\infty} (-1)^n \int dx
 \Bigl( f( q^{ 2n +1 } x ) + f( q^{-(2n+1)} x ) \Bigr) .
\ee
The last equality in (\ref{1.13}), showing the explicit ($q\leftrightarrow
q^{-1}$)-symmetry, follows due to identity
 \[2(T_q+T_{q^{-1}})^{-1}=T_{q^{-1}}(1+T_{q^{-2}})^{-1}+T_q(1+T_{q^{2}})^{-1} .
 \]

 Applying S-TD derivative and S-TD integral to $x^n $ yields
\be \label{1.14}
 D_x x^n = \{ n \} x^{n-1},  \qquad \int Dx x^n = \frac{x^{n+1}}{\{ n+1
\} } ,
\ee
\bd
 \int D x \frac{1}{x}  = \frac{2}{q +\qi} \ln x , ~~~
D_x \left( \frac{2}{q +\qi} \ln x \right) = \frac{1}{x} ,
 \ed
and also
\[
D_x^k x^n = \frac{\{ n \}!}{\{ n-k\}! } x^{n-k} .
\]

 The S-TD-exponential function ${\cal E} (x) $ is
 defined as\footnote{Note that,
 unlike well-known $q$-exponents $e_q(x)$ and $E_q(x)$ (given e.g. in \cite{Gasper}),
 the ${\cal E} (x)$ possesses straightforward $q\to 1$ limit to the usual exponent.}
\be \label{1.15}
{\cal E} (x) = \sum_{n=0}^{\infty} \frac{1}{\{ n \}!} x^n,
\ee
 or
\be \label{1.16}
 {\cal E} (x) = 1 + \sum_{n=1}^{\infty} \frac{ 2^n q^{ \frac{1}{2}
n(n-1) } }{n! \prod_{k=0}^{n-1} ( 1 + (q^2 )^k ) } x^n .
\ee
 It is worth to note that, whereas the TD-type $q$-exponent
  for which the convergence of the series analogous to (\ref{1.15}),
  with respect to ordinary $\exp (x)$, improves at $q>1$ and worsens at $q<1$,
  the convergence of the series in (\ref{1.15}) - (\ref{1.16})
  is better for any positive $q$.

 One can verify that the inverse $[{\cal E}(x)]^{-1}$ differs from ${\cal E}(-x)$.
 Let us give the ${\cal E}^{-1}(x)$ explicitly, namely
  \be \label{1.17}
\frac{1}{{\cal E}(x)}=\sum_{n=0}^{\infty}b_n x^n
\ee
 where $b_0=1$, and all the other expansion coefficients
 are found using recursion relation
\[
b_n=-\frac{1}{\{n\}}-\sum_{j=1}^{n-1}b_j\frac{1}{\{n-j\}!} , \qquad
n\ge 1 ,
 \]
 inferred from the equality:
  $ \frac{1}{{\cal E}(x)} \sum_{n=0}^{\infty}\frac{x^n}{\{n\}!}=1$.
 Let us write first few coefficients:
\[
  b_0=1, \quad \ b_1=-1,
 \quad \
b_2=-\frac{1}{\{2\}!}+1, \quad \
b_3=-\frac{1}{\{3\}!}+\frac{2}{\{2\}!}-1,
\]
\[
 b_4=-\frac{1}{\{4\}!}+\frac{2}{\{3\}!}-\frac{3}{\{2\}!}+\frac{1}{(\{2\}!)^2}+1,
\]
\[
 b_5=-\frac{1}{\{5\}!}+\frac{2}{\{4\}!}-\frac{3}{\{3\}!}+\frac{4}{\{2\}!}
     +\frac{2}{\{3\}!\{2\}!} -\frac{3}{(\{2\}!)^2}-1.
 \]

 The S-TD $q$-exponent ${\cal E} (x)$ can be also presented in yet
 another form. Namely,
\be \label{1.19}
 {\cal E} (x) = \sum_{n=0}^{\infty} \frac{q^{\frac{1}{2}n(n-1)}}
 {n!(-1, q^2 )_n }  (2x)^n
 \ee
 where the notation $(a,Q)_n$ means:
\[(a,Q)_n\equiv (1-a)(1-aQ)(1-aQ^2)\cdots (1-aQ^{n-1}), \qquad
(a,Q)_0=1.
\]

 It is easy to check that the S-TD-exponential function satisfies
\be \label{1.18}
D_x {\cal E} ( a x ) = a {\cal E} ( ax ),
\qquad \int Dx {\cal E} ( ax ) = \frac{1}{a} {\cal E} ( ax )
\ee
 for an arbitrary constant $a$.
 From the relation (assume $a > 0$)
\be \label{1.19}
\int_0^{\infty} Dx {\cal E} ( -a x ) =\frac{1}{a}, \hspace{10mm}
\ee
we can obtain the following formula
\be \label{1.20}
 \int_0^{\infty} Dx {\cal E} ( -a x
) x^n = \frac{(-1)^n }{a^{n+1}}\prod_{k=1}^n \{ - k \}  .
\ee
 Inserting $a=1$ into eq.(\ref{1.20}) yields
\be \label{1.21}
\int_0^{\infty} Dx {\cal E} ( - x ) x^n = (-1)^n
 \prod_{k=1}^n \{ - k \}.
\ee
 The latter relation can be rewritten in the form
\be \label{1.22}
\int_0^{\infty} Dx {\cal E} ( - x ) x^n =\frac{ 2 \{ n+2\} !}{ \{ 2 \}
( n+1) ( n+2 ) }
\ee
 if one uses the formula
\[
\{- k \} = - \frac{ k}{k+2}\{ k+2 \}.
\]

As last point in this section, let us show that the S-TD exponential
given in (\ref{1.17}) can be presented as some bibasic
hypergeometric function \cite{Gasper} whose general definition is
\[ _{r,r'}\mathcal{F}_{s,s'}
  (\underline a_r;\underline c_{r'}|\underline b_s;\underline
 d_{s'};(q,p);z)\equiv\]
 \[\equiv\sum_{n=0}^{\infty}{\frac{(a_1,a_2,\mbox{...},a_r;q)_{n}
 (c_1,c_2,\mbox{...},c_{r'};p)_{n}}{(b_1,b_2,\mbox{...},b_s;q)_{n}
 (d_1,d_2,\mbox{...},d_{s'};p)_{n}\,(q;q)_n}}\times\]
\[
\times\left[ (-1)^{n} q^{n(n-1)/2}  \right]^{1+s-r}  \left[ (-1)^{n}
p^{n(n-1)/2}  \right]^{s'-r'}\! z^{n}.
\]
Here $\underline{a}_m\!=\!(a_1,a_2,\mbox{...},a_m)$
and
$(a_1,a_2,\mbox{...},a_m;q)_{n}\equiv\prod_{k=1}^m(a_k,q)_n$;
likewise,
$(c_1,c_2,\mbox{...},c_{m'};p)_{n}\equiv\prod_{l=1}^{m'}(c_l,p)_n$.
Indeed, taking into account that
\[(a^2,q^2)_n=(a,q)_n(-a,q)_n, \qquad
  \lim_{p\to 1} \frac{(p,p)_n}{(1-p)^n}= n! \,
\]
 from (\ref{1.17}) we obtain the desired result:
    \vspace{-1mm}
\begin{equation} \label{1.25}
        { \cal E }(x) \equiv\exp^{\rm (S-TD)}_q(z)\equiv\
        \lim_{p\rightarrow1}\
        _{0,1}\mathcal{F}_{0,2}(-;0| -;{\rm i},-{\rm i};(p,q);(1-p)z).
\end{equation}
Thus, like in the case on non-symmetric TD $q$-exponential
\[  \exp^{\rm(TD)}_q = \sum_{n=o}^{\infty} \frac{ q^{-\frac{1}{2}n(n-1)}}
 {n!} x^n
\]
considered in \cite{GKL-13}, the symmetric S-TD type $q$-exponential
${ \cal E } (x)$ studied in this paper also takes the form of some
(limit of) bibasic hypergeometric function.

\section{Non-Fibonacci (Quasi-Fibonacci) nature of S-TD oscillator}
 Diverse deformed oscillator models appeared in quantum physics till present time
\cite{A-C76,Biedenharn,Macfarlane89,Chakrabarti91,Odaka,Chaturvedi,GR07},
\cite{Arik-92,Chung-93,GKR-10}. All of them have different structure
functions and different properties. Since some of them where known
as Fibonacci ones \cite{Arik-92}, the question arose about some
classification of deformed oscillator models. As a criterion for
possible classification of DOs, in \cite{GKR-10} we proposed to
check the validity (or failure) of the Fibonacci-like recurrence
relation
 \be \label{2.1}
 E_{n+1}=\lambda E_n+\rho
E_{n-1},    \qquad   n\geq 1 ,     \qquad
E_n=\frac12(\varphi_{n+1}+\varphi_{n}) ,
 \ee
 for each three consecutive energy levels, with {\it constant} real
 $\lambda$ and $\rho$ (recall that $\varphi_{n}$ is the structure function
 of deformation).
 Note that usual quantum harmonic oscillator is a Fibonacci one: it
 satisfies eq. (\ref{2.1})
 with $\lambda=2$ and $\rho$=-1.

In this section we will check whether symmetric Tamm-Dancoff
$q$-deformed oscillator defined in (\ref{1.1}) belongs to the class
of Fibonacci oscillators. To do that, we substitute the
corresponding expressions for the energy of S-TD deformed
oscillator in (\ref{2.1}):
\[ (n+1)(q^n+q^{-n})+(n+2)(q^{n+1}+q^{-n-1})=\]
\[=\lambda\Bigl(n(q^{n-1}+q^{-n+1})+(n+1)(q^{n}+q^{-n})\Bigr)+\]
\[
+\rho\Bigl((n-1)(q^{n-2}+q^{-n+2})+n(q^{n-1}+q^{-n+1}\Bigr) .
\]
 From this, considering independent basis elements,
 we infer the following set of equations:
  \[
nq^n:  \quad \ 1+q=\lambda(q^{-1}+1)+ \rho (q^{-2}+q^{-1}); \]
\be
\label{2.3}
 q^n: \quad \ \ \ 1+2q=\lambda-\rho q^{-2};
\ee
\[
nq^{-n}: \quad 1+q^{-1}=(q+1)(\lambda+\rho q);
\]
\[
q^{-n}:  \quad \ 1+2q^{-1}=\lambda-\rho q^{2} .
\]
The first two equations are solved to give $\lambda=2q$ and
$\rho=-q^2$. These values of $\lambda$ and $\rho$, however, cannot
satisfy third and fourth equations if $q\ne 1$. Therefore we
conclude that the system of equations (\ref{2.3}) is incompatible and
the symmetric Tamm-Dancoff oscillator does not belong to the class
of Fibonacci oscillators.

Following the ideas of \cite{GKR-10}, we consider for the S-TD oscillator the
{\it quasi-Fibonacci} recurrence relation that involves the
coefficients $\lambda=\lambda(n)\equiv\lambda_n$ and
$\rho=\rho(n)\equiv\rho_n$: \be \label{2.4} E_{n+1}=\lambda_n
E_n+\rho_n E_{n-1}, \qquad n\geq 1. \ee
 To certify the quasi-Fibonacci nature of the S-TD oscillator we have
to find $\lambda_n$ and $\rho_n$ explicitly.
 There exist different ways to find $\lambda_n$ and $\rho_n$,
 and here we follow two of them.

\underline{First way (``simplest splitting'')}. Replace the
quasi-Fibonacci relation (\ref{2.4}) by the system of two equations
for $\lambda_n$ and $\rho_n$ in terms of the structure function
$\varphi(n)$, see (\ref{2.1}):
\begin{equation} \label{2.5}
\cases{ \varphi(n+1)=\lambda_n \varphi(n)+\rho_n \varphi(n-1), \cr
\varphi(n+2)=\lambda_n \varphi(n+1)+\rho_n\varphi(n).}
\end{equation}
From these we find \be \label{2.6}
\lambda_n=\frac{\varphi(n+1)-\rho_n\,\varphi(n-1)}{\varphi(n)} ,
\qquad
\rho_n=\frac{\varphi(n+2)\varphi(n)-\varphi^2(n+1)}{\varphi^2(n)
-\varphi(n+1)\varphi(n-1)}. \ee
 It is clear (for more details see \cite{GKR-10}) that each
 non-Fibonacci deformed oscillator can be viewed as quasi-Fibonacci
one, the coefficients $\lambda_n$ and $\rho_n$ being dependent on
the particular model through the choice of structure function
$\varphi(n)$.

 Recalling explicit structure function $\varphi_{\tiny{STD}}(n)$ of S-TD
deformed oscillator given in (\ref{1.4}), we finally obtain
 \be \label{2.7}
\lambda_n=\frac{2(1+q^{-2n})(q+q^{2n-1})-(n-1)(n+2)
(q+q^{-1})(q-q^{-1})^2}{(1+q^{-2n})(q^2+q^{2n-2})-n^2(q-q^{-1})^2},
\ee
 \be  \label{2.70}
\rho_n=-\, \frac{(q^n+q^{-n})^2 -
n(n+2)(q+q^{-1})^2}{(1+q^{-2n})(q^2+q^{2n-2})-n^2(q-q^{-1})^2}.
\ee
 Note that in the limit $q\to 1$ we have $\lambda_n\to 2$ and
$\rho_n\to -1$ as for the usual quantum oscillator.
 It is easy to verify that since  the coefficients $\lambda_n$ and $\rho_n$
 in  (\ref{2.6}) -- (\ref{2.70}) are the solutions of the system of equations (\ref{2.5}),
 they obviously satisfy the quasi-Fibonacci relation (\ref{2.4}).

Following \underline{second, most general, way} (see \cite{GKR-10}) for
deriving the
coefficients $\lambda_n$ and $\rho_n$,  
we consider the system
 \be \label{2.8}
 \cases{ E_{n+1}=\lambda_n E_n+\rho_n E_{n-1}, \cr
 E_{n+2}=\lambda_{n+1} E_{n+1}+\rho_{n+1}
E_n,}
\ee
 which cannot be solved uniquely (high degree of
arbitrariness). Therefore we take the first relation in (\ref{2.8}) and
use the explicit structure function of S-TD oscillator:
  \[
(n+1)(q^n+q^{-n})+(n+2)(q^{n+1}+q^{-n-1})=\]
\[=
\lambda_n(n(q^{n-1}+q^{-n+1})+(n+1)(q^n+q^{-n}))+
\]
\be \label{2.9}
 +\rho_n((n-1)(q^{n-2}+q^{-n+2})+n(q^{n-1}+q^{-n+1})).
\ee
 The latter equation can be replaced equivalently by the two equations
\be \label{2.10}
0=(\lambda_n-2)(n+1)(q^n+q^{-n})-K(q,n)(n+1)(q^n+q^{-n}) ,
\ee
\[
(n+2)(q^{n+1}+q^{-n-1})-(n+1)(q^n+q^{-n})=
 (\lambda_n+\rho_n) n (q^{n-1}+q^{-n+1})+
\]
 \be\label{2.100}
+\rho_n(n-1)(q^{n-2}+q^{-n+2})+K(q,n)(n+1)(q^n+q^{-n}) \ee
 (adding (\ref{2.10}) to (\ref{2.100}) recovers (\ref{2.9})).
 Here $K(q,n)$ is an arbitrary function incorporating the arbitrariness
 of splitting eq. (\ref{2.9}) into two equations, and such that
 $K(q,n)\to 0$ when $q\to 1$.
  Now, from the equation (\ref{2.10}) one easily finds $\lambda$:
  \be\label{2.11}
  \lambda_n=2+K(q,n),
 K(q,n)\, {\buildrel {\scriptstyle q\to 1}\over \longrightarrow}\ 0 .
   \ee
  Substituting this $\lambda_n$ in (\ref{2.10}),
  we obtain the coefficient $\rho_n$, namely
\[ \rho_n=\frac{(n+2)(q^{n+1}+q^{-n-1}) -
(n+1)(q^n+q^{-n})-2n(q^{n-1}+q^{-n+1})}{n(q^{n-1}+q^{-n+1})+(n-1)(q^{n-2}+q^{-n+2})} -
\]
\be\label{2.12}
\hspace{6mm}  - \frac{K(q,n)\Bigl(n(q^{n-1}+q^{-n+1})+(n+1)(q^n+q^{-n})\Bigr)}
 {n(q^{n-1}+q^{-n+1})+(n-1)(q^{n-2}+q^{-n+2})}. \ee
Note that at $q\to 1$ we obtain $\lambda_n\to 2$,\ $\rho_n\to -1$ as
it should be (the usual oscillator case).

It is useful to check \underline{two special cases} of $K(q,n)$:

1. If $K(q,n)=0$, we have

\ \ \ \ $\lambda_n=2$,\ and
 \[\hspace{-11mm} \rho_n= \frac{(n+2)(q^{n+1}+q^{-n-1})-(n+1)(q^n+q^{-n})-2n(q^{n-1}+q^{-n+1})}
 {n(q^{n-1}+q^{-n+1})+(n-1)(q^{n-2}+q^{-n+2})}  \ \
 {\buildrel {\scriptstyle q\to 1}\over \longrightarrow}\ -1 .
 \]

2. On the other hand, it is possible that
\[\hspace{-11mm}\rho_n = -1 \quad {\rm if} \quad K(q,n)=-1+\widetilde{K}(q,n), \qquad {\rm where}\]

\[\hspace{-11mm}
 \widetilde{K}(q,n) =
 \frac{(n+2)(q^{n+1}+q^{-n-1})+(n-1)(q^{n-2}+q^{-n+2})}
  {(n+1)(q^{n}+q^{-n})+n(q^{n-1}+q^{-n+1}) }  .
\]
\hspace{13mm} {Then,
$\lambda_n=1+ \widetilde{K}(q,n)$.
At $q\to 1$, we have $\widetilde{K}(q,n)\to 1$, \ $\lambda_n\to 2$.
Remark

\hspace{5mm} that both in case 1 and case 2, one coefficient is still $n$-dependent.}

\section{Relating S-TD oscillator with $p,q$-deformed oscillators}
 It is an interesting question whether the
one-parameter symmetric Tamm-Dancoff can be ``embedded'' in the
two-parameter deformed oscillator. Consider the five-parameter
oscillator whose structure
function is build by ``gluing'' (using the weight-like $t$-parameter,
 see formula (22) in \cite{GKL-13}) the two different 
structure functions (here -- two different copies of
$p,q$-oscillator):
\be\label{3.1}
\widetilde{\varphi}_{p,q,P,Q;t}=t
\varphi_{p,q}+(1-t)\varphi_{P,Q}, \quad {\rm where}  \ee
\[
 \varphi_{p,q}=\frac{p^N-q^N}{p-q},
\qquad \varphi_{P,Q}=\frac{P^{N}-Q^{N}}{P-Q}.
\]
Recall that $p,q$-deformed oscillator with structure function
 $\varphi_{p,q}$ is a Fibonacci one \cite{Arik-92}.
  The same is true for the $P,Q$-deformed oscillator defined by $\varphi_{P,Q}$.
  But, that is not true in general for the hybrid deformed oscillator given by
  $\widetilde{\varphi}_{p,q,P,Q;t}$ (see the Proposition below).

Let us put $P=p^{-1}$,\ $Q=q^{-1}$, and $t=\frac12$. That yields
 \be\label{3.2}
\widetilde{\varphi}_{p,q}=
 \frac12\left(\varphi_{p,q}+\varphi_{p^{-1},q^{-1}}\right) ,
\ee
 and it is evident that in the special case of $p=q$,
 this two-parameter deformed oscillator yields nothing but
 the (structure function of) S-TD
 deformed oscillator.

 {\bf Proposition.} The five-parameter $(t;p,q,P,Q)$-deformed
 oscillator given by the structure function $\widetilde{\varphi}_{p,q,P,Q;t}$
 in (\ref{3.1}) is not Fibonacci (except for the two cases:
 $p=P$, $q=Q$ and $P=q^{-1}$, $Q=p^{-1}$).
Thus, it should be viewed as quasi-Fibonacci one.

The proof proceeds similarly to that based on (\ref{2.3}), but now the
analogous system of 4 equations stems from equating the coefficient
expressions (involving $q,p,Q,P,\lambda,\rho$) at $p^n$, $q^n$,
$P^n$ and $Q^n$ viewed as independent basis elements.

{\bf Remark.}  An alternative viewpoint can be also adopted:
deformed oscillator with the structure function (\ref{1.1}) can be
viewed as two-dimensional or two-mode deformed oscillator whose two
modes are (i) independent, and (ii) each one is described by a copy
of mutually dual (in the sense of $q\to q^{-1}$, $p\to p^{-1}$
duality) $p,q$-oscillators and likewise, at $p=q$, by a copy of
 two mutually ($q\leftrightarrow q^{-1}$)-dual TD oscillators.

\section{Pairwise energy levels degeneracy of S-TD oscillator}
 Taking the Hamiltonian as
 \[
 H=\frac12 (a^\dagger a + a a^\dagger),
 \]
we have its eigenvalues
 \[
 H |n \rangle = E_n |n \rangle
 \]
 where
\be  \label{4.3}
 E_n =\frac14 [(n+1)(q^{n}+q^{-n})+ n
(q^{n-1}+q^{1-n})]
\ee
(note that $E_0=\frac12$, as for the usual
non-deformed quantum oscillator).

To study possible accidental degeneracy of, say, neighboring levels,
let us consider
 \[ \Delta E_n = E_{n+1} - E_n=
\]
\[ \ = \frac14 [(n+2)(q^{n+1}+q^{-n-1})- n (q^{n-1}+q^{-n+1})]=
\]
\be  \label{4.4}   = \frac14 [2(q^{n+1}+q^{-n-1})
 +n(q-q^{-1})(q^n-q^{-n})] > 0 .
 \ee
 The latter follows for $q>0$  due to the inequality $q+q^{-1}\geq 2$
 (saturated at $q=1$), and we conclude that there is no degeneracy
 for real positive values of $q$.

The situation however \underline{differs if $q$ is phase-like}.
 So, let us assume
 \[ q=\exp({\rm i}\theta) , \hspace{20mm}  -\pi \leq \theta \leq \pi .
 \]
 In this case, from (\ref{4.4}) we have:
 \[   E_{n+1} - E_n=
 \frac12[(n+2)\cos((n+1)\theta)- n \cos((n-1)\theta)] =
 \]
 \[
= \cos(n\theta)\cos\theta - (n+1)\sin(n\theta) \sin\theta =
\]
\[
=\frac12 [(n+2)T_{n+1}(x) - n T_{n-1}(x) ] = \]
\[   = (n+1)T_{n+1}(x) - n x T_n(x) , \hspace{20mm}  x\equiv\cos\theta
, \]
 where $T_n(\cos\theta)$ is the Chebyshev polynomial of
 $n$-th order.

It is possible to solve $E_{n+1} - E_n=0$ at any fixed $n$.
 As instances, consider few cases:  $n=0,1,2,3$.
\[\hspace{-25mm}
 \underline{E_1=E_0}. \qquad  \textrm{This holds at} \ \theta = \theta_0 =\pm
 \frac{\pi}{2}, \quad {\rm or} \quad q=\pm{\rm i}.
\]
\[\hspace{-25mm}
 \underline{E_2=E_1}. \qquad \textrm{This holds at} \ \theta = \theta_1=\pm
\arcsin \frac{\sqrt3}{3}.
\]
\[\hspace{-25mm}
\underline{E_3=E_2}. \qquad \textrm{This holds at} \ \theta =
\theta_2=\pm \arcsin \frac{\sqrt2}{4} \ (\textrm{and also at}\
\theta = \theta_0).
\]
\[\hspace{-25mm}
\underline{E_4=E_3}. \qquad  \textrm{This holds at}\ \theta =
\theta_3 =\pm \arcsin \sqrt{\frac{17\pm\sqrt{209}}{40}}.
\]
 In a similar fashion one can prove that other cases of pairwise
 degeneracy may occur.
 Namely, there exist
 special value(s) of the  parameter $q=\exp ({{\rm i}\theta})$
 that solve the equation
\[
E_{n+r}=E_n, \ \ \  n\geq 0, \ \ \  r>1 .
\]
The latter can be presented as
\[
(n+r+1)(q^{n+r}\!+q^{-n-r}) + (n+r)(q^{n+r-1}\!+q^{-n-r+1}) -\]
\[
- (n+1)(q^{n}+q^{-n})- n(q^{n-1}\!+q^{-n+1}) = 0
\]
or, in terms of Chebyshev polynomials, as
\[\hspace{-20mm}
 (n+r+1)T_{n+r}(x) + (n+r)T_{n+r-1}(x)\!-\!(n+1)T_n(x)\!-\!n T_{n-1}(x) = 0 ,
  \quad   x\equiv\cos\theta ,
\]
or as
\[\hspace{-20mm}
(n+1)\Bigl( T_{n+r}(x)-T_n(x) \Bigr) + n \Bigl( T_{n+r-1}(x) -
T_{n-1}(x) \Bigr) + r \Bigl( T_{n+r}(x)+ T_{n+r-1}(x) \Bigr) = 0 .
\]
For instance of solving, we quote two cases:
\[\hspace{-25mm}
 \underline{E_2=E_0}. \qquad \textrm{This holds at} \quad \theta = -\pi \quad \textrm{or at} \quad
 \theta=\arccos \frac23.
 \]
\[\hspace{-25mm}
 \underline{E_3=E_1}. \qquad \textrm{This holds at} \quad \theta = -\pi \quad \textrm{or
at} \quad \theta=\arccos \frac{5\pm\sqrt{89}}{16}.
\]

\section{ Coherent states}
In this section, we construct the coherent states of the
STD-oscillator  algebra (\ref{1.1})-(\ref{1.2}). The coherent state
$|z\rangle$ is defined as an eigenstate of the annihilation operator
in the form
 \be \label{5.1}
 a |z\rangle = z |z\rangle .
 \ee
 The coherent state can be also represented by using the Fock
 eigenvectors of the number operator:
 \be \label{5.2}
  |z \rangle = \sum_{n=0}^{\infty} c_{n}
(z) |n\rangle.
\ee
 Inserting  eq.(\ref{5.2}) into eq.(\ref{5.1}), we have
\be \label{5.3}
\sum_{n=1}^{\infty } c_{n} (z) \sqrt{ \{ n \} }
|n-1\rangle = \sum_{n=0}^{\infty } zc_{n} (z) |n\rangle.
\ee
From eq.(\ref{5.3}), we get the following recurrence relation
\be \label{5.4}
c_{ n+1} = \frac{z}{\sqrt{ \{ n+1 \} } } c_n
  ~~~ (n=0, 1, 2, \cdots ).
\ee
Solving this relation (\ref{5.4}), we infer
\[
c_n = \frac { 1} {\sqrt{ \{ n \}!} }z^n  c_0 .
\]
From $\langle z |z\rangle =1 $ it is not difficult to obtain \be
\label{5.6} c_{0}^{-2} = { \cal E } ( |z|^2 ). \ee
 Thus the coherent state takes the form
\be \label{5.7}
|z \rangle = \frac{1}{\sqrt{{ \cal E } (|z|^2)}} \sum_{n=0}^{\infty}
\frac {z^n }{\sqrt{ \{ n \}!} }  |n\rangle .
\ee

 Now let us show that the coherent state $|z\rangle $ forms a
 complete set of states. To establish this, we
invoke the completeness relation:
\be \label{5.8}
\frac{1}{ \pi } \int \int
|z\rangle \mu( |z|^2 ) \langle z | |z| D|z| d \theta = I,
\ee
 where $ \mu ( |z|^2 ) $ is a weight function.
  Inserting eq.(\ref{5.7}) in eq.(\ref{5.8}), we obtain
\be \label{5.9}
\sum_{n=0}^{\infty} \frac{ 1}{ \{ n \}! }|n\rangle
\langle n | \int_0^{\infty} f(x)  x^n Dx = I
\ee
 where
$ x = |z|^2 $ and $ f(x) = {\mu (x)}/{{ \cal E } (x )}$.
 If we find $f(x) $  satisfying
\be \label{5.10} \int_0^{\infty} f(x)  x^n D x =  \{ n \}! \ , \ee
the completeness is proved.
 It is useful to set
\be \label{5.11} f(x) = { \cal E } ( -x ) g(x) = { \cal E } (-x) \sum_{k=0}^{\infty}
g_k x^k ,\ee
that implies ${\mu (x)} = {{ \cal E } (x )} { \cal E } ( -x ) g(x) $.

 Inserting eq.(\ref{5.11}) into eq.(\ref{5.9}), with the account of (\ref{1.22})
 we obtain
\be   \label{5.12}
 \sum_{k=0}^{\infty} g_k \frac{ ( n+k)!}{n!} \prod_{j=n+1}^{n+k+2} \phi(j) = \phi (2),  \qquad
 \phi(j) \equiv \frac{ \{ j \} }{ j }  .
 \ee
  From  (\ref{5.12}) we infer the recurrence relation:
  \bd g_k = \frac{ (-1)^k }{k! \phi ( k+2
) ! } \left[ \phi (2) - \sum_{i=0}^{k-1} g_i \frac{ (-1)^i k! }{ (
k-i )! } \prod_{j= k - i +1 }^{ k+2 } \phi (j) \right], ~~(k \ge 1 )
\ed
 \bd  
 g_0 =1 , ~~\phi(i)! = \prod_{j=1}^i \phi(j) \ed
One easily gets the particular coefficients $g_k$:
 \bd    g_0 =1 ,  \qquad
  g_1 = \frac{ \phi (2) ( \phi (3) - 1 ) }{\phi(3)! } ,
 \ed
 \bd   
g_2 = \frac{ \phi (2) - \phi(3) \phi(4) + 2 \phi (2) ( \phi (3) - 1
) \phi(4) }{2!\phi(4)! }, \ed
 and so on. This shows in a constructive way the existence of the
 function $g(x)$ and, through ${\mu (x)} = {{ \cal E } (x )} { \cal E } ( -x ) g(x) $,
 of the weight function $\mu(|z|^2)$.
 Thus, the completeness is proved.

\section{Conclusions}
 Within the symmetric $q$-deformed Tamm-Dancoff
oscillator proposed in this paper and possessing the symmetry
$q\leftrightarrow q^{-1}$ we explored a number of its aspects. We
constructed both the Fock type representation, and the coordinate
space realization.
 For the needs of the latter we introduced and studied in some detail the
corresponding non-standard version of $q$-calculus, first of all
 the S-TD type $q$-differentiation and $q$-integration.
 The related S-TD type $q$-exponential function given in (\ref{1.15}) or
 (\ref{1.17}), possesses interesting properties (e.g. under $q$-integration),
 as shown in (\ref{1.17}) and (\ref{1.18}) -- (\ref{1.22}).
  Besides the inverse (\ref{1.17}), we have found its presentation as a particular case,
  see (\ref{1.25}), of  bibasic hypergeometric function.

Among the special properties possessed by S-TD $q$-oscillator, we
establish its quasi-Fibonacci (thus non-Fibonacci) nature and
 the ability to present the S-TD $q$-oscillator as a special case
 of some family of $p,q$-deformed oscillators.
  Let us emphasize that the  result concerning quasi-Fibonacci nature
  of S-TD $q$-oscillator  (see Sec.~3 and especially Sec.~4) suggests the
  simple way of how to build many new quasi-Fibonacci oscillators taking
  two (or more) Fibonacci ones as building blocks.
 Because of $q\leftrightarrow q^{-1}$-symmetry, the studied $q$-oscillator
 admits not only real, but also phase-like deformation parameter and, due to
 this fact, the possibility of pairwise accidental energy level degeneracy arises.
 At last, we provide the S-TD type $q$-coherent states and prove for these
 the completeness relation.

 It would be interesting to study, first, the connection
 of the S-TD $q$-oscillator with respective deformation of Heisenberg
 algebra for the position and momentum operators like in \cite{Chung-klimyk}
 and, second, the thermostatistics of S-TD type $q$-deformed Bose gas model
 (linked with the S-TD $q$-oscillators).

  Also it is of interest to find for 
  the proposed S-TD $q$-oscillators, with complex $q$ enabled by
  $q\leftrightarrow q^{-1}$ symmetry, useful 
  physical applications.
  To argue that the use of complex, phase-like form of deformation parameter $q$
  may open new possibilities, let us mention some known applications
  of $q$-deformed  algebras.
  In the first one, based on quantum algebras replacing the Lie algebras
  of flavor groups $SU(n)$, the $q$-parameter $q=\exp({\rm i}\theta)$
  can be linked \cite{G-NP}     
  with the Cabibbo angle $\theta_c$ of quark flavor mixing.
  Next, this same angle $\theta_c$ may be of relevance when a $q$-Bose gas model
  with phase-like $q$ 
  is used to effectively describe \cite{G-NP}   existing data on the
  correlation intercepts of pions generated in relativistic heavy-ion collisions.
  In the third application, concerning 2+1 quantum gravity, the approach of J.Nelson
 and T.Regge leads \cite{NR-97} to the nonstandard $q$-algebras
 $U'_q(so_n)$, see \cite{G-99}, with the deformation parameter
 linked to cosmological constant $\Lambda$ as
 $q=\frac{4\alpha -{\rm i}h}{4\alpha +{\rm i}h}$ where $\alpha^2=-\frac{1}{3 \Lambda}$,
 $\Lambda < 0$, and $h$ being Planck constant.
 At last, within the $p,q$-deformed Bose gas model involving
 mutually conjugate complex $p, q$, the deformed critical temperature $T_c^{p,q}$
 was found to rise \cite{GR-12} with growing extent of deformation.

\ack{ This work was partly supported by the Special Program of the Division of
 Physics and Astronomy of the NAS of Ukraine (A.M.G.) and by the Grant (A.P.R.)
  for Young Scientists of the NAS of Ukraine (No.~0113U004910).}

\end{document}